# Conceptual Understanding of Computer Program Execution: Application to C++

**Sabah Al-Fedaghi**

**Computer Engineering Department, Kuwait University**
**P.O. Box 5969 Safat 13060 Kuwait**

## Abstract

A visual programming language uses pictorial tools such as diagrams to represent its structural units and control stream. It is useful for enhancing understanding, maintenance, verification, testing, and parallelism. This paper proposes a diagrammatic methodology that produces a conceptual representation of instructions for programming source codes. Without loss of generality in the potential for using the methodology in a wider range of applications, this paper focuses on using these diagrams in teaching of C++ programming. C++ programming constructs are represented in the proposed method in order to show that it can provide a foundation for understanding the behavior of running programs. Applying the method to actual C++ classes demonstrates that it improves understanding of the activities in the computer system corresponding to a C++ program.

## 1. Introduction

This paper aims at proposing a diagrammatic methodology that produces a conceptual representation of instructions for programming source codes. Without loss of generality, this methodology is applied to the programming language C++. The advantages of this application include source code understanding, maintenance, verification, testing, and identification of parallelism, in addition to other purposes such as reuse and reverse engineering. The proposed methodology enhances *understanding* of program code in terms of its corresponding computer operations, not as registers, fetch/store/execute cycle, addresses; rather, in terms of *conceptual* operations such as create, release, transfer, receive, and process, thus completing the cycle of understanding, where it is always claimed that, as a first step, a programmer must understand the application domain (e.g., inventory).

This work can be considered a type of visualization of computer programs. Program visualization is a well-known paradigm. A visual programming language, not to be confused with a visual programming environment, is a language that uses graphic tools to represent structural units and control streams in programs. This type of language facilitates creating and specifying of program elements graphically rather than by writing them textually [1]. Some visualization of programs is based on the notion of dataflow programming that represents a program as a directed graph of the data flowing among operations [2][3].

Program understanding is one of the most important aspects influencing the maintainability of programs for programmers [4]. "Mechanisms for improving program comprehensibility can reduce maintenance cost and maximize return on investments in legacy code by promoting reuse" [5]. According to Kiper et al. [5],

> The entire software engineering philosophy is built on the premise that high level language code is created for human consumption rather than driven by machine requirements. The first step in repairing or modifying existing code is to understand what that code does.

Nevertheless,

> Complete understanding of a large system is an unrealistic goal. Rather, a maintainer must identify those program components that are important for a specific change and focus on understanding them well enough to safely make the modification. It is hard to define exactly how programmers go about achieving this level of understanding or even how they know when it has been achieved. [6]

Some visual programming languages express constructs in diagrams. Diagrams are often used in software learning and development, and in business systems to represent requirements, dataflows, workflows, and software architecture [7]. For years, diagrams have been utilized in constructing software systems [8][9]. Many tools have been built to aid programmers, giving them many capabilities, including drawing and sketching to construct programs and examine codes [10][11][12].

> Diagrams - or more generally, visualizations of non-apparent systems, concepts, relationships, processes and ideas - help students to recognise and *understand* parallels and structural correlations between things in the world; their constitutive natures, their internal structures and relationships; the systems of which they form a part, and the processes they are involved with. [13] [Italics added]

According to Lee [14],

> Despite all of these previous efforts, the majority of programming activity occurs in text-centric development environments with information often conveyed through list and tree views. If programmers worked efficiently and effectively in these environments, there may be little reason to consider how to better support programming through diagramming tools.

Here, it can be sensed that the need exists to develop tools to facilitate understanding and to serve more than documentation and initial planning needs of a program, as in the case of pseudo codes and flowcharts.

This paper claims that: *A new methodology of high-level description, called the Flowthing Model (FM), is a viable alternative to other diagrammatic methods for program understanding.*
To substantiate this claim, programming constructs in a textbook will be recast in FM, with the aim of showing the advantages and disadvantages of each method.

To focus such a process, the paper narrows the materials as follows:
- The paper focuses mainly on the general problem of using diagrams as a "foundation for *understanding the behaviour of running programs*" in learning programming [15].
- Without loss of generality of the potential for using FM in a wider range of application, this paper focuses on using diagrams in *teaching* programming [15].
- More focus is realized in the paper by taking the construct of C++ as a study case to exemplify and contrast FM with flowcharting and pseudo codes.
This tightening of materials in the paper makes it easier to concentrate on a limited domain, thus achieving the capability to explore specific aspects of the proposed method.

## 2. Problem

The skill of programming is quite valuable, and interest in programming is increasing; however, there are difficulties in learning to program [15]. It is reported that a novice needs about 10 years of practice to become an expert programmer [16].

> Acquiring and developing knowledge about programming is a highly complex process. It involves a variety of cognitive activities, and mental representations related to program design, program understanding, modifying, debugging (and documenting). Even at the level of computer literacy, it requires construction of conceptual knowledge, and the structuring of basic operations (such as loops, conditional statements, etc.) into schemas and plans. [17]

Computer science students have problems in mastering programming. According to Thomas [18], this difficulty is "one of the manifestations of lack of understanding of program behavior."

> In order to understand a program's behaviour it is necessary for the programmer to have a model of the computer that will execute it. This 'notional machine' provides a foundation for understanding the behaviour of running programs… Programming ability must rest on a foundation of *knowledge about computers*, a programming language or languages, … [15].

Typically, models of program comprehension (e.g., [19] [20]) concentrate on programming that involves mappings from the problem domain into the programming domain. In these approaches, there is little appreciation of the role at the *computation level* (Fig. 1). High-level programming languages are supposed to be abstracted from machine hardware. Meanings of architectural aspects such as the CPU, address, Memory, ALU, … are brushed off in chapter 1 of most programming texts. The reason is to avoid getting involved in computer hardware but to focus on the software instead as the tool for problem solving.

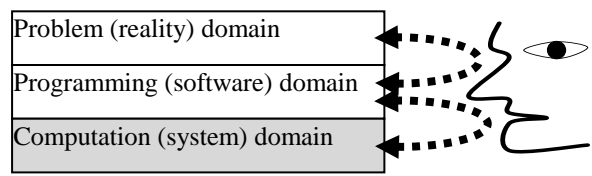

Fig. 1 Different domains related to programming

This paper demonstrates that it is possible to build a conceptual map of activities corresponding to those in a C++ program within a computer system without incorporating hardware elements. Here the term *conceptual* refers to a high-level depiction of essential elements and their interrelationships in the computation domain (computer) using diagrammatic notations. Its purpose is to convey a common description without technological aspects that can serve as a guide for understanding operations specified in a C++ program.

Nevertheless, the generality of FM applications at different levels of programming development (see Fig. 2), can be claimed. For example, a simplified FM conceptual description introduces a more complete flowchart. In addition, a textual specification of the FM depiction is suggested as a narrative of events that is less "sketchy" than pseudo code.

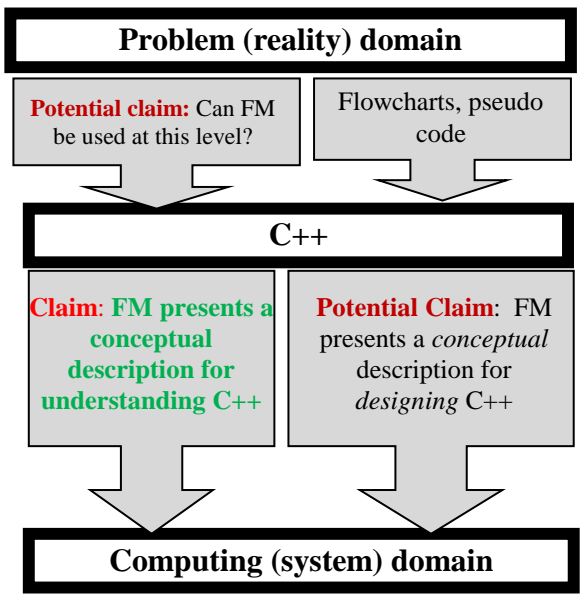

Fig. 2 Domains and claims in the paper

For the sake of a self-contained paper, section 3 briefly describes the Flowthing Model, FM, upon which the new representation is built. FM has been utilized in many applications [e.g., 21–25].

## 3. Flowthing Model

The Flowthing Model (FM) is a depiction of the structure of a system, a road map of its components and conceptual flow. A *component* comprises *spheres* (e.g., operating system, program, statement, C++ function) that may enclose or intersect with other spheres (e.g., the sphere of a house contains rooms, which in turn include walls, ceilings). Or, a sphere embeds flows (called *flowsystems* - e.g., walls encompass pipes of water flow and wires of electricity flow).

Things that flow in a flowsystem are referred to as *flowthings* (e.g., money, data, products, cars, parts). The life cycle of a flowthing can be defined in terms of six mutually exclusive *stages*: creation, process, arrival, acceptance, release, and transfer. Within a certain sphere:
- *Creation* means the appearance of the flowthing in the totality of a sphere's system for the first time (e.g., the creation of a new program).
- *Process* means application of a change to the form of an existing flowthing (e.g., writing a program in a structured way).
- *Release* means marking a flowthing as "to be output", but it remains within the sphere (e.g., data marked "to be transmitted").
- *Transfer* denotes the input/output module of the sphere (e.g., interface component [port] of a device for a communication channel).
- *Arrival* means that the flowthing reaches the sphere but is not necessarily permitted to enter it (e.g., a letter delivered to the wrong recipient and rejected, to be returned).
- Acceptance means permitting the arrived flowthing to enter the system.
Fig. 3 shows a flowsystem with its stages, where it is assumed that no released flowthing flows back to previous stages. The reflexive arrow in the figure indicates flow to the Transfer stage of another flowsystem. For simplicity's sake, the stages Arrive and Accept can be combined and termed *Receive*.

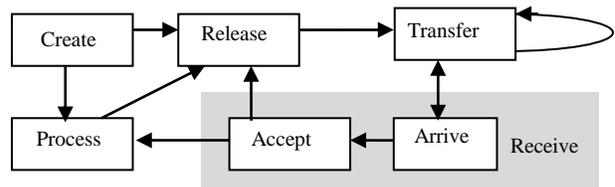

Fig. 3 Flowsystem

The *stages* in the life cycle of a flowthing are mutually exclusive (i.e., the flowthing can be in one and only one stage at a time). All other states of flowthings are not generic states. For example, we can have stored created flowthings, stored processed flowthings, stored received flowthings, etc. Flowthings can be released but not transferred (e.g., the channel is down), or arrived but not accepted, …
In addition to flows, *triggering* is a transformation (denoted by a dashed arrow) from one flow to another, e.g., a flow of electricity triggers the flow of air.

**Example**: This example is artificially constructed to somewhat resemble a C++ program with two statements, one to fetch data from computer memory to be displayed on the screen, and the second to signal that such an operation is successful. The purpose is to illustrate the FM concepts of sphere, flowsystem, and flowthing in a computer program. Note that when a sphere includes a single flowsystem, one rectangle is drawn to represent both of them, the sphere as well as its flowsystem.

In a market, the daily procedure for display of gold and precious jewelry, under the supervision of a manager, is performed by a worker who performs the following two tasks in sequence, as shown in an FM representation in Fig. 4:

Task 1: Bring the jewels from the safe to be exhibited.
Task 2: Report to the manager the success of the opening operation.

Accordingly, the market sphere includes all other subspheres. The first task involves three subspheres: safety box, worker, and exhibition. It starts when the gold and jewels (a flowsystem - circle 1 in the figure) *flow* from the sphere of the safety box (2) to the worker, then to the exhibition area (3). In the exhibition sphere, the jewels are unpacked from their boxes (processed) and displayed (4). The second task, reporting (a sphere/flowsystem, 5) is accomplished by creating an "OK" message (a flowthing, 6) and sending it to the manager (7).

The conceptual picture involves two flows: that of the gold and jewels, and that of information (OK message). Triggering can be added to the figure; say, the manager triggers the worker to start setting up the daily exhibit. It is assumed that the two tasks are executed in sequence; otherwise, it is possible to make the end of task 1 trigger task 2.

## 4. Conceptual Base for Understanding C++

This section presents the main contribution of this paper: FM-based description of C++ constructs. The representation depicts the conceptual (in contrast to hardware) computer operations that correspond to these constructs. The course *CpE-200: Computer Programming for Engineers* is selected for an experiment with FM modeling of programming. It uses the text *C++: How to Program*, Fifth Edition by Deitel and Deitel (Prentice Hall, 2005) as the source of the sample programs in this paper after removing comments. The course objectives in the academic catalog are stated as follows:
- Familiarize the students with fundamental *understanding of computers* and the basic constructs of a modern programming language. [Italics added]

- Familiarize the student with the basic problem-solving concepts, top-down design, stepwise refinement, modularity, object oriented programming, and reusability .

### 4.1 Input *cin* and output *cout*

Starting with the semantics of *cin* in C++, Fig. 5(a) shows its FM representation. Data *flows* from the keyboard, assuming standard input/output, to the data flowsystem in the computer sphere, to be stored. Because in this conceptual picture there is only one type of flowthing (data), it is possible to depict it with one rectangle, as shown in Fig. 5(b). Fig. 5(c) shows *cout*, and Fig. 5(d) shows a sample output to the screen.

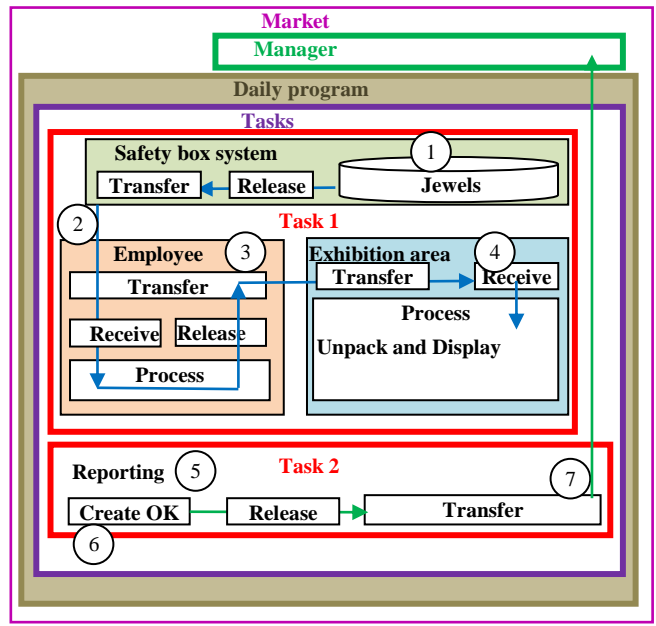

Fig. 4 FM description of jewel exhibition process

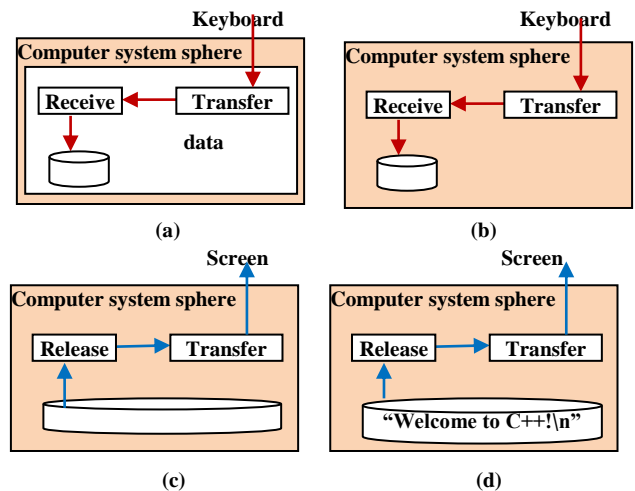

Fig. 5 FM description of *cin* and *cout*

The FM representation reveals conceptual incompleteness of flows, as seen in the previous figures, where keyboard and screen seem to be something outside the picture. Going beyond C++, it is possible to represent the flow of data in *cout* to the screen sphere, as shown in Fig. 6 for the screen. The flow from the keyboard in *cin* is shown in Fig. 7. In Fig. 7, the user's actions trigger (dashed arrow) the generation of data that flows to the computer system. Note that in FM, an **action** is a flowthing that can be created, released, transferred, received, and processed.

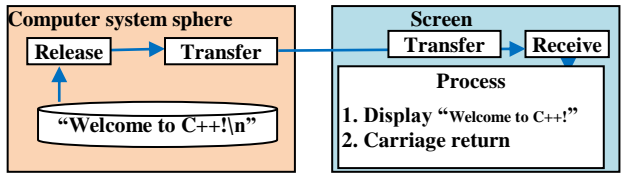

Fig. 6 FM description of *cout* and Screen

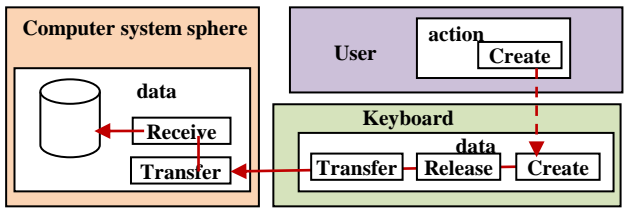

Fig. 7 User action triggers generating of data in the keyboard that flows to the computer system

## 4.2 Program 1

Now consider the C++ program 1 shown in Fig. 8. For comparison, Fig. 9 shows its FM representation. The process starts at circle 1, where the first statement is executed by triggering *cout* (circle 2) to retrieve the string (3) from memory that flows to the screen (4). Then the *return* statement is executed (5), and (if execution is successful) it creates zero (6) that flows to the operating system (7).

Notice the resemblance of this program to the example given in Fig. 4 in the previous section. Main, statements, operating system, memory system, and screen as well as Market, manager, employee, and exhibition are all conceptual spheres.

In the Computer sphere, there are the sub-spheres Operating system, and Main. Main includes the Statements sphere. In this sphere there are statement 1 and statement 2. Statement 1 has three flowsystems: Memory, *cout* and screen. The second statement has the flowsystem of the return signal. Since statement 2 has a single flowsystem, they are represented by one rectangle.

Fig. 9 presents a conceptual picture of operations without discussion of the computer hardware. The FM representation provides a complete conceptual description of the process of execution inside the computer. The execution is controlled by the operating system sphere that activates the program main sphere statements in sequential order. In the statement sphere, the execution starts with the first statement, then the second statement.

This computer-based view of the program enhances its understanding. The statement:

   *std::cout << "Welcome to C++!\n";*

presents some mysteries to the student: What is *"Welcome to C++!\n"*? Where was it? What is *cout*? Is it the screen? Where is the screen? Usually, the answers are given in textual or oral format, but now FM presents a blueprint of this statement just like the blueprint for water and electrical connections in a building. Of course such a map enhances understanding more than the analogous architectural description: *pipes << water*. In the CpE-200 class, students agreed with this conclusion; however, they complained about the complexity of the representation.

```
1    int main ()
2    {
3       std::cout << "Welcome to C++!\n";
4       return 0
5    }
```

Fig. 8 C++ program 1

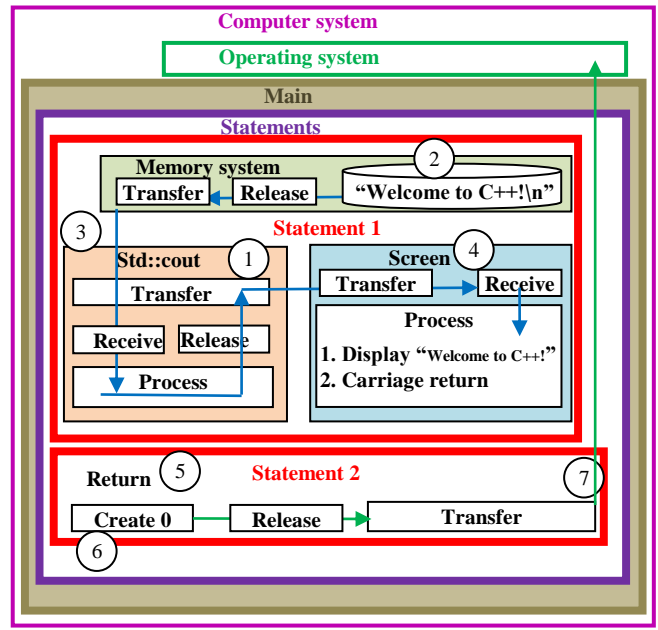

Fig. 9  FM description of program 1

### 4.3 Program 2

Now consider the program shown in Fig. 10 and its corresponding FM representation in Fig. 11, which shows that within the main sphere, in the statements sphere, statements are executed in sequence. Statements
*std::cin >> integer1;*
*std::cin >> integer2;*
can be drawn in one box in Fig. 11.

```
1     int main()
2     {
3         int integer1;
4         int integer2;
5         int sum;
6         std::cin >> integer1;
7         std::cin >> integer2;
8         sum = integer1 + integer2;
9         std::cout << "Sum is " << sum << std::endl;
10        return 0;
11    }
```

Fig. 10 C++ program 2

The execution starts by waiting for user input (1) through the keyboard (2) that is received by *cin* (3) to be sent to the memory system (4) and stored in location *integer1* (5). Similarly, flowsystems (6–8) deposit *integer2* in the memory. In the statement *Sum = integer1 + integer2;* the values of *integer1* and *integer2* (9–10) flow to the ALU (11), where addition is performed (12) to trigger generating (13) the result that flows to the memory (14). As in program 1, the statements are executed sequentially and return creates zero and sends it to the operating system.

The rest of statements in Fig. 11 can be explained in a similar fashion to similar constructs discussed previously. Sub-statements
*std::cout << "Sum is "*
*std::cout << sum << std::endl;*
in
*std::cout << "Sum is " << sum << std::endl;*
can be drawn in a single box in Fig. 11.

Notice how the assignment statement is represented. Understanding this modeling would, certainly, eliminate any confusion between the semantics of the statement and the meaning of the symbol "=" in the statement. Variables are also clearly defined in terms of name, value, and type. *Integer1*, *integer2*, and *sum* are names of locations in the memory. This is usually repeated by the teacher, but it hardly "sticks" in the students' minds the way a picture depicted by the FM representation does. Different types of variables are emphasized by flowing in different flowsystems.

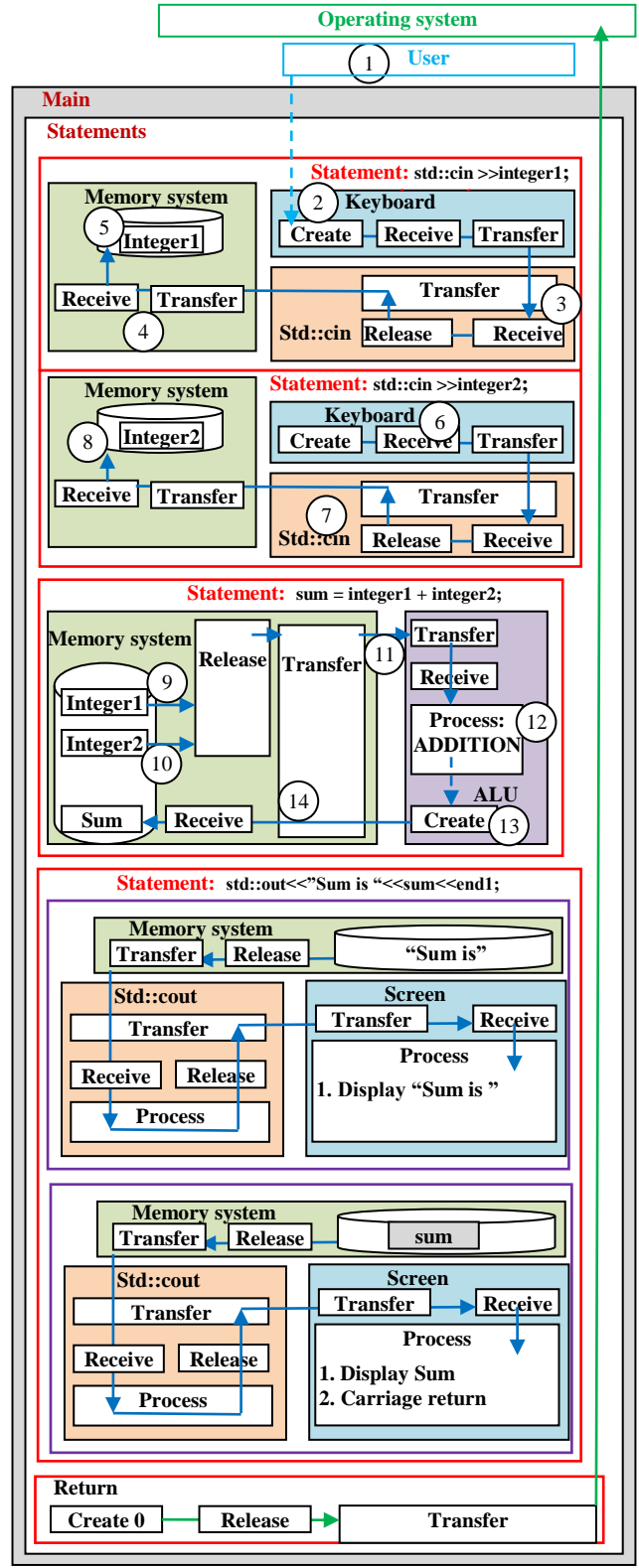

Fig. 11 FM description of *cout* and Screen

## 4.4 *If* statement

Fig. 12 shows a C++ program that involves an *if* statement. Fig. 13 shows its FM representation. The computing process starts with the flow of "Welcome to C++!\n" and "the relationships they satisfy:" (1) to the screen (2). Then *num1* and *num2* are input through the keyboard (3) to flow and be stored in the memory, as described previously (4).

```
1   int main ()
2   {
3     int num1;
4     int num2;
5     cout << "Enter two integers\n"
6
7     cin >> num1 >> num2;
8     if (num1 == num2)
9        cout << num1 << " is equal to " << num2 << endl;

      Other if statements
40     return 0;
42   }
```

Fig. 12 C++ program 3

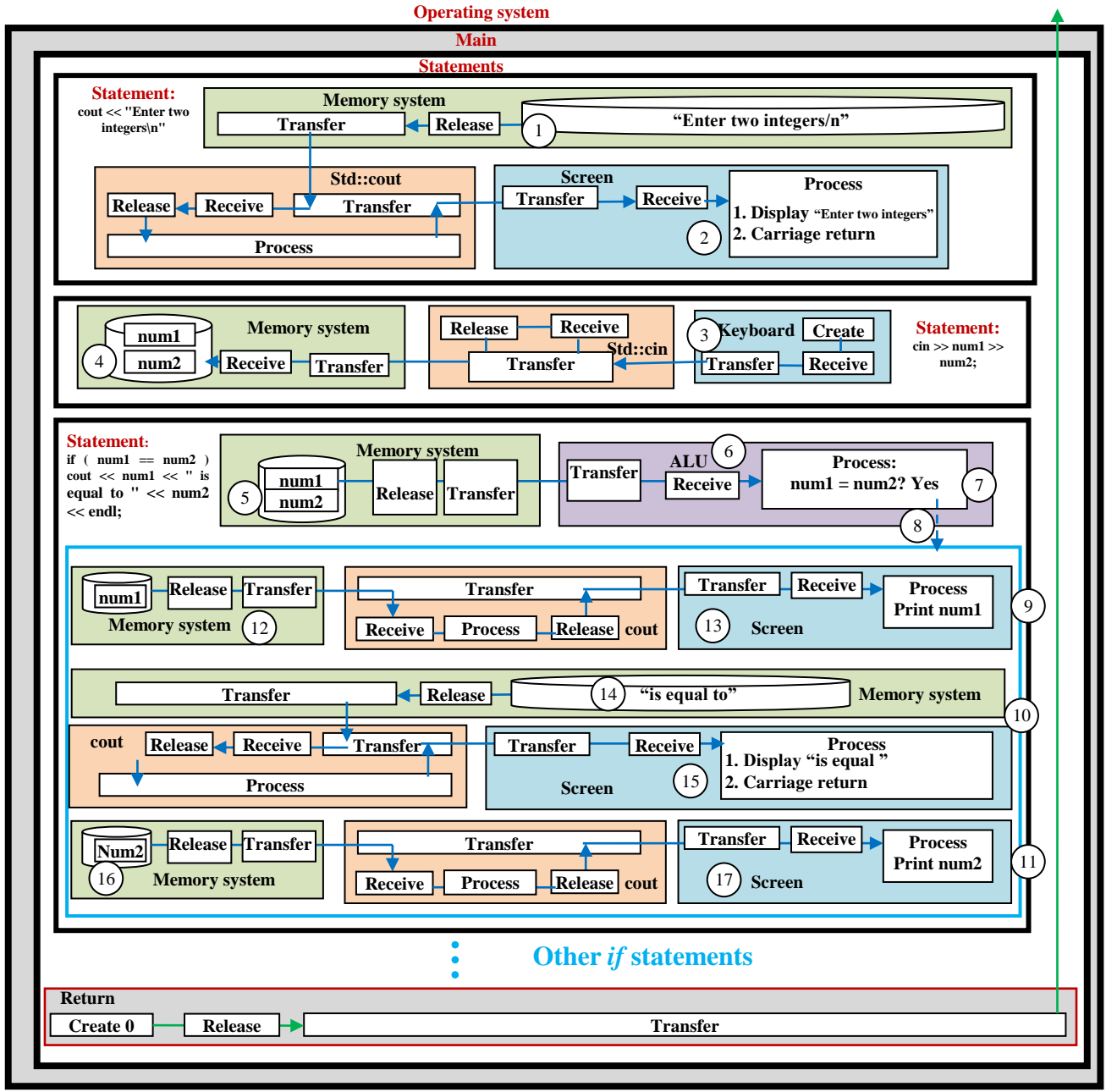

Fig. 13 FM representation of C++ program 3

The *if* statement is executed by retrieving *num1* and *num2* from memory (5) to flow to the ALU (6). In the ALU, the two integers are compared, and if they are equal (7), then this triggers (8) three output constructs (9–11). In output 9, the value of *num1* (12) flows to the screen (13). In output 10, the string " is equal to " (14) flows to the screen (15). In output 11, the value of *num2* (16) flows to the screen (17).

Figure 14 shows a simplified version of the FM representation, where a diamond is used for an *if* statement in the fashion of flowcharts. Fig. 15(a) shows further simplification to arrive at the flowchart shown in Fig 15(b). Accordingly, the FM representation provides a complete description of the process that is sketched by flowcharts.

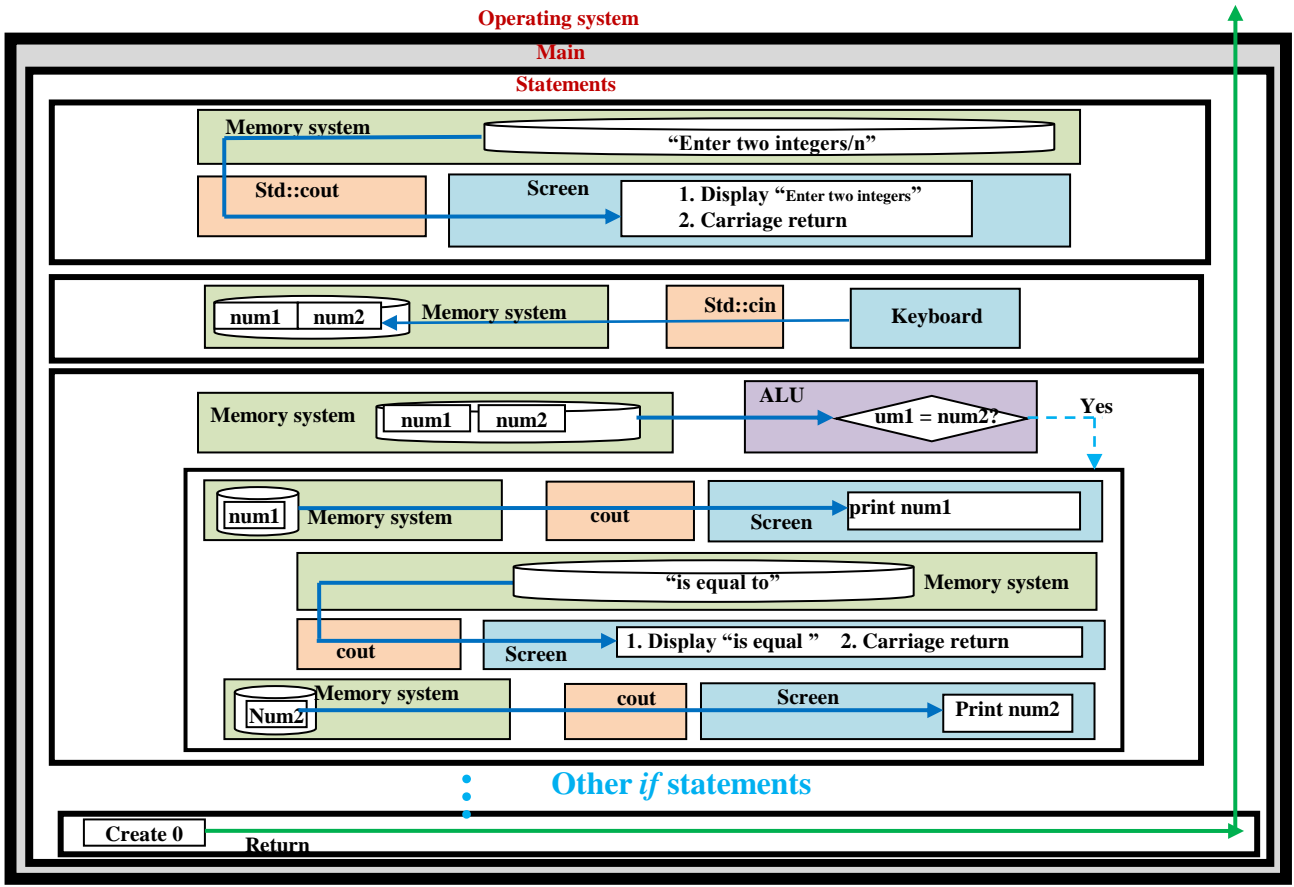

Fig. 14 Simplified version of FM representation of C++ program 3

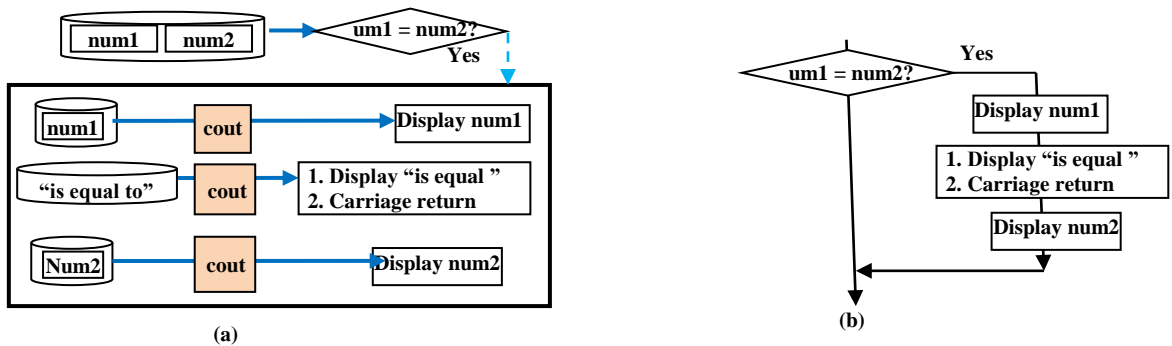

Fig. 15 From FM to flowchart

## 4.5 *While* statement

Fig. 16 shows a portion of the C++ program that involves a *while* statement, and Fig. 17 depicts its FM representation. The program problem can be stated as follows:

*A class of ten students took a quiz. The grades (integers in the range 0 to 100) for this quiz are available to you. Determine the class average on the quiz.*

The execution of the While statement starts at circle 1, where the value of gradecounter is sent to the ALU to be compared to 10. For simplicity's sake, we ignore here the issue that the constant 10 itself is fetched from memory. Accordingly, if gradecounter ≤ 10, the block in the brackets { } is executed (3). Four tasks are triggered:

A.  "Enter grade" is printed (4).

B.  *grade* is input (5)

C.  The values of *grade* and *total* are added and the result is stored in total (6)

D.  *gradecounte*r is incremented by 1 (7)

Figure 18 shows a simplified version of the while loop.

```
22    while (gradeCounter <= 10) {
23        cout << "Enter grade: ";
24        cin >> grade;
25        total = total + grade;
26        gradeCounter = gradeCounter + 1;
27    }
```

Fig. 16 While statement

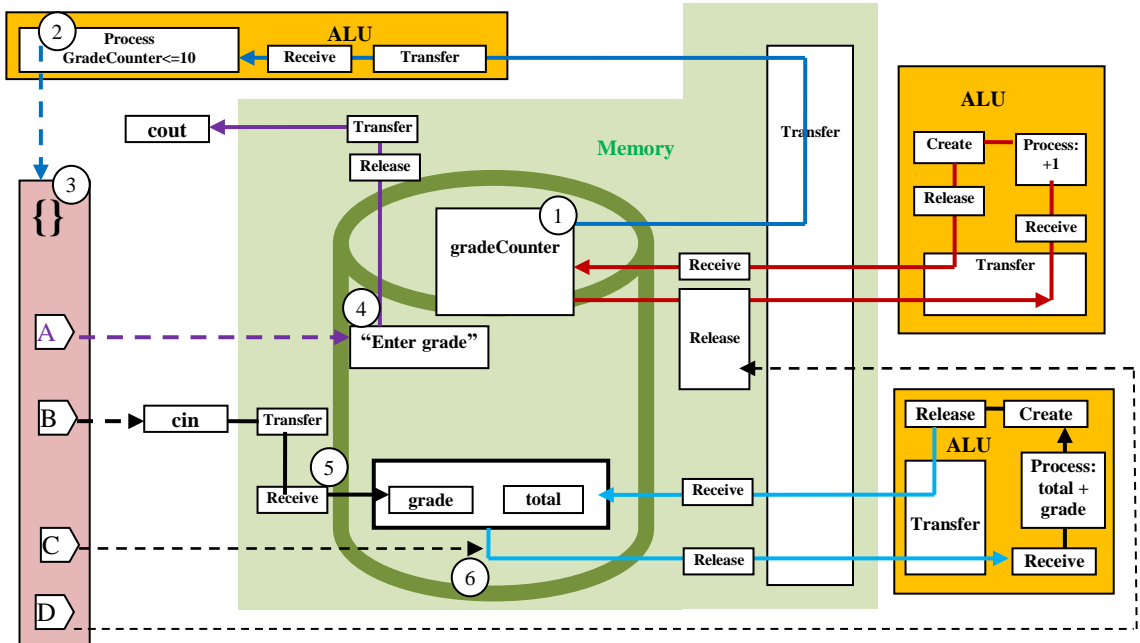

Fig. 17 FM representation of C++ program 3

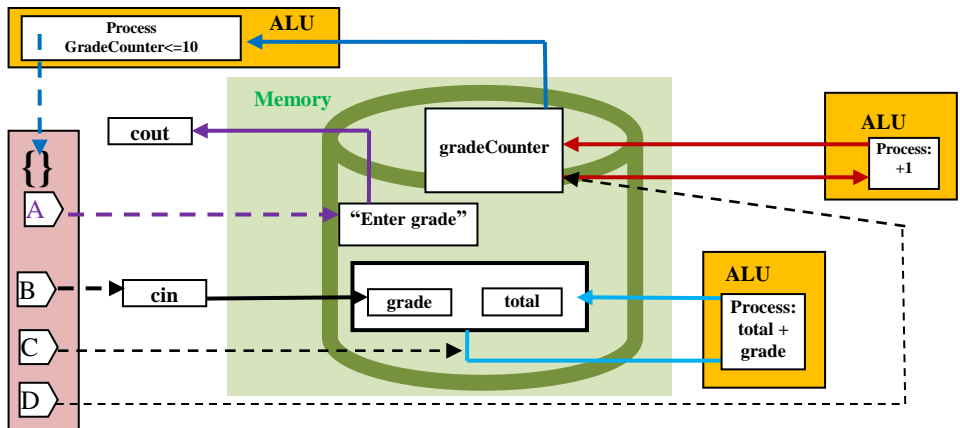

Fig. 18 Simplified FM representation of *While* statement

## 5. Conclusions

This paper proposes a diagrammatic methodology that produces a conceptual representation of instructions for programming source codes. The paper focuses on diagrams used for teaching C++ programming. C++ programming constructs are represented in the proposed method in order to show that it can provide a foundation for understanding the behavior of running programs. The paper introduces the methodology for the purpose of facilitating discussion about the FM model, and to report initial findings in its application.

The method is being applied in a yearlong study to explore its potential uses. The initial results are as follows:

- Some students complained at the beginning that the FM method is complex. The instructor then showed them design diagrams from different engineering design application (blueprints of buildings, electrical systems, aerodynamics, etc.). The argument is in order to build a precise specification of a system, then, when it seems complicated diagrams are necessary, as long as they are developed in a systematic way. FM has few concepts that are repeatedly applied in different parts of the schemata. Programs, especially those embedded in critical systems (e.g., heart control instruments, airplanes) ought to be fully understood and specified.

- Students have indicated that their understanding of C++ increased when the instructor explained the semantics of the language utilizing the FM model. It should be pointed out that this method is utilized side by side with the typical (oral) explanation of C++ statements.

- Early indicators point to the fact that the FM methodology benefits analysis of programs, but not as a method to construct them as in the case of pseudo code. Nevertheless, since designing and building programs is an iterative process, some students reported that the FM method helped in rewriting their programs after they'd written earlier versions and examined their FM semantics.

- As demonstrated in Fig. 15, FM representation can be simplified and reduced to flowcharts. This gives more meaning to the origin of flowcharts based on conceptual operations inside the computer. Similar results can be applied to pseudo codes.

We can conclude that the FM method as applied in this paper presents a new viable approach in the programming domain; however, its advantages/disadvantages are still to be explored in two areas:

- Experimentation with the method in actual programming environments.
- Development of a friendly user interface for FM, with possible auto-diagramming of programs and statements [26].

In the current experiment, information about exam results is being collected over the course of two semesters, and results will be reported within a year.

## References

[1] S. D. Bragg, and C. G. Driskill, "Diagrammatic-Graphical Programming Languages and DoD-STD-2167A", IEEE Systems Readiness Technology Conference, "Cost Effective Support into the Next Century", Conference Proceedings, 1994, pp. 211–220.

[2] W. M. Johnston, J. R. P. Hanna, and R. J. Millar, "Advances in Dataflow Programming Languages", ACM Computing Surveys, Vol. 36, No. 1, 2004, pp. 1–34.

[3] Akepogu Anand Rao, and Karanam Madhavi, "Framework for visualizing model-driven software evolution and its application", IJCSI International Journal of Computer Science Issues, Vol. 7, Issue 1, No. 3, January 2010.

[4] Victor R. Basili, "Viewing Maintenance as Reuse-Oriented Software Development", IEEE Software, Vol. 7, No. 1, 1990.

[5] James Kiper, Chuck Ames, and Lizz Howard, "Using Program Visualization to Enhance Maintainability and Promote Reuse", in Proceedings of Computing in Aerospace, March 1995, American Institute of Aeronautics and Astronautics.

[6] Norman Wilde, Understanding Program Dependencies, Carnegie Mellon University, Software Engineering Institute, SEI-CM-26, August 1990.

[7] Dhirendra Pandey, Ugrasen Suman, and A. K. Ramani, "A Framework for Modelling Software Requirements", IJCSI International Journal of Computer Science Issues, Vol. 8, Issue 3, No. 1, May 2011.

[8] A. Von Mayrhauser, and A. M. Vans, "Program Comprehension during Software Maintenance and Evolution", Computer, Vol. 28, No. 8, 1995, pp. 44-55.

[9] M. Cherubini, G. Venolia, R. DeLine, and A. J. Ko, "Let's Go to the Whiteboard: How and Why Software Developers Draw Code", in ACM Conference on Human Factors in Computing Systems (CHI), 2007, pp. 557–566.

[10] J. Grundy, and J. Hosking, "Supporting Generic Sketching-Based Input of Diagrams in a Domain-Specific Visual Language Meta-Tool", in 29th International Conference on Software Engineering (ICSE), 2007, pp. 282–291.

[11] V. Sinha, D. Karger, and R. Miller, "Relo: Helping Users Manage Context during Interactive Exploratory Visualization of Large Codebases," in IEEE Symposium on Visual Languages and Human-Centric Computing (VL/HCC'06), 2006, pp. 187–194.

[12] M. A. D. Storey, and H. A. Müller, "Manipulating and Documenting Software Structures Using SHriMP Views", in 11th International Conference on Software Maintenance (ICSM), 1995, pp. 275.

[13] John Cussans, "Diagram As Thinking Machine", DRUGG Presentation, Art as Metapractice (Part One). http://diagramresearch.wordpress.com/symposia/

[14] Seonah Lee, Gail C. Murphy, Thomas Fritz, and Meghan Allen, "How Can Diagramming Tools Help Support Programming Activities?" in IEEE Symposium on Visual Languages and Human-Centric Computing (VL/HCC 2008), Herrsching am Ammersee, Germany, September 15–19, 2008.

[15] A. Robins, J. Rountree, and N. Rountree, "Learning and Teaching Programming: A Review and Discussion", Computer Science Education, Vol. 13, No. 2, 2003.

[16] L. E. Winslow, "Programming Pedagogy – A Psychological Overview", SIGCSE Bulletin, Vol. 28, 1996, pp. 17–22.

[17] J. Rogalski, and R. Samurçay, "Acquisition of Programming Knowledge and Skills", In J. M. Hoc, T.R.G. Green, R. Samurcay, and D. J. Gillmore (Eds.), Psychology of Programming, pp. 157–174. London: Academic Press, 1990.

[18] Lynda Thomas, Mark Ratcliffe, and Benjy J. Thomasson, "Can Object (Instance) Diagrams Help First Year Students Understand Program Behaviour?" Diagrams, 2004, pp. 368–371.

[19] R. E. Brooks, "Towards a Theory of the Cognitive Processes in Computer Programming", International Journal of Man-Machine Studies, Vol. 9, 1977, pp. 737–751.

[20] F. P. Brooks Jr., The Mythical Man-Month: Essays on Software Engineering, Anniversary Edition. Reading, MA: Addison-Wesley, 1995.

[21] Sabah Al-Fedaghi, "Reconceptualization of Class-based Representation in UML", IJCSI International Journal of Computer Science Issues, Vol. 9, Issue 6, 2012.

[22] Sabah Al-Fedaghi, "Diagrammatization of the Transmission Control Protocol", IJCSI International Journal of Computer Science Issues, Vol. 9, Issue 5, 2012.

[23] Sabah Al-Fedaghi, "Scrutinizing the Rule: Privacy Realization in HIPAA", International Journal of Healthcare Information Systems and Informatics (IJHISI), Vol. 3, No. 2, 2008, pp. 32-47.

[24] Sabah Al-Fedaghi, "Scrutinizing UML Activity Diagrams", 17th International Conference on Information Systems Development (ISD2008), Paphos, Cyprus, August 25–27, 2008.

[25] Sabah Al-Fedaghi, "Information Management and Valuation", International Journal of Engineering Business Management, Vol. 4, 2012, p. 47.

[26] Soumaya Amdouni, Wahiba Ben Abdessalem Karaa, and Sondes Bouabid, "Semantic Annotation of Requirements for Automatic UML Class Diagram Generation", IJCSI International Journal of Computer Science Issues, Vol. 8, Issue 3, No. 1, May 2011.

**Sabah Al-Fedaghi** holds an MS and a PhD in computer science from the Department of Electrical Engineering and Computer Science, Northwestern University, Evanston, Illinois, and a BS in Engineering Science from Arizona State University, Tempe. He has published two books and more than 70 articles published/forthcoming in more than 50 different peer-reviewed journals. He has also published more than 80 articles in conferences on Software Engineering, Database Systems, Information Systems, Computer/information Ethics, Information Privacy, Information Security and Warfare, Conceptual Modeling, and Artificial Agents. He is an associate professor in the Computer Engineering Department, Kuwait University. He previously worked as a programmer at the Kuwait Oil Company and headed the Electrical and Computer Engineering Department (1991–1994) and the Computer Engineering Department (2000–2007).